\RequirePackage{fix-cm}
\documentclass[twocolumn,epjc3]{svjour3}  
\smartqed  
\RequirePackage{graphicx}
\journalname{Eur. Phys. J. C}
\usepackage{cite} 
\usepackage{amsmath}
\usepackage{hyperref}
\usepackage{cite}
\usepackage{amsmath,amssymb}
\usepackage{color}

\usepackage{float}
\usepackage{graphicx}
\usepackage{bbm}
\usepackage{graphicx}
\usepackage{amssymb}
\usepackage{epstopdf}
\usepackage{color}
\usepackage{ulem}
\usepackage{graphicx}
\usepackage{pdfpages}

\begin{document}
\title{{Class I polytropes for anisotropic matter.} }


\author{A. Ramos  \thanksref{e1,
        addr1} 
        \and
        C. Arias \thanksref{e2,addr1}
        \and
        E. Fuenmayor \thanksref{e3,addr2}
        \and
        E. Contreras\thanksref{e4,addr3}
}
\thankstext{e1}{e-mail: 
\href{mailto:anthony.ramos@yachaytech.edu.ec}{\nolinkurl{anthony.ramos@yachaytech.edu.ec}} }
\thankstext{e2}{e-mail: 
\href{mailto:cynthia.arias@yachaytech.edu.ec}{\nolinkurl{cynthia.arias@yachaytech.edu.ec}} }
\thankstext{e3}{e-mail: 
\href{mailto:ernesto.fuenmayor@ciens.ucv.ve}{\nolinkurl{ernesto.fuenmayor@ciens.ucv.ve}}}
\thankstext{e4}{e-mail: 
\href{mailto:econtreras@usfq.edu.ec}{\nolinkurl{econtreras@usfq.edu.ec}}}
\institute{School of Physical Sciences \& Nanotechnology, Yachay Tech University, 100119
Urcuqu\'{i}, Ecuador\\ \label{addr1}
\and
Centro de F\'isica Te\'orica y Computacional, Facultad de Ciencias, Universidad Central de Venezuela\label{addr2}
\and
Departamento de F\'isica, Colegio de Ciencias e Ingenier\'ia, Universidad San Francisco de Quito,  Quito, Ecuador\label{addr3}
}

\date{Received: date / Accepted: date}

\maketitle

\begin{abstract}
In this work we study class I interior solutions supported by anisotropic polytropes. The generalized Lane--Emden equation compatible with the embedding condition is obtained and solved for a different set of parameters in both the isothermal and non--isothermal regimes. For completeness, the Tolman mass is computed and analysed to some extend. As a complementary study we consider the impact of the Karmarkar condition on the mass and the Tolman mass functions respectively. Comparison with other results in literature are discussed.
\end{abstract}

\section{Introduction}\label{intro}
Polytropic equations of state have played a remarkable role in astrophysics (see \cite{1pol,3pol} and references therein), and have been extensively  used to study the stellar structure. For example, white dwarfs have been modeled considering  Newtonian polytropes \cite{abellan20,LHN} while in
more compact configurations (e.g. neutron stars, super Chandrasekhar white dwarfs) \cite{herrera2013,prnh} general relativistic
polytropes have been extensively studied (see, for example, and the references therein \cite{4a, 4b, 4c, 5, 6, 7, 8, 9, 10}).

For an isotropic fluid, the theory of polytropes is based on the polytropic equation of state which reads
\begin{equation}
P=K\rho_0^{\gamma}=K\rho_{0}^{1+1/n} ,\label{Pola}
\end{equation}
where $P$ denotes the isotropic pressure,
$\rho_{0}$ stands for the mass (baryonic) density and $K$, $\gamma$, and $n$ are usually called  the polytropic constant, polytropic exponent, and polytropic index, respectively. Once the equation of state (\ref{Pola}) is assumed, the whole system is described by the Lane--Emden equation that may be numerically solved for any set of the parameters of the theory. When the constant $K$ is calculated from natural constants, the polytropic equation of state may be used to model a completely degenerate Fermi gas in the non-relativistic ($n = 5/3$) and relativistic limit ($n = 4/3$). In this case, Eq. (\ref{Pola}) provides
a way of modeling compact objects such as white dwarfs and allows   to obtain in a rather direct way the Chandrasekhar mass limit. Otherwise, if $K$ is a free parameter, the models can be used to describe either an isothermal ideal gas or a completely convective star. These models related to isothermal ideal gas are relevant in the so-called Sch\"onberg--Chandrasekhar limit \cite{3pol}.

Although local isotropy is a very common assumption in the study of compact objects, there is strong evidence that suggests that for certain ranges of density, a large number of physical phenomena can cause local anisotropy. A possible source is related to intense magnetic fields observed in compact objects such as white dwarfs, neutron stars or magnetized strange quark stars \cite{11,12,13,14,15}. Another source is the high indexes of viscosity expected to be present in neutron stars, in highly dense matter produced by opacity of matter to neutrinos in the collapse of compact objects \cite{16,17,18}, and in the superposition of two isotropic fluids, to name a few. It is important to note that, although the degree of anisotropy may be small, the effects produced on compact stellar objects may be appreciable \cite{herrera97,pr10,Ovalle:2016pwp,Ovalle:2018vmg,ovalle17,ovalle18,ovalle19}. So, the assumption of an isotropic pressure is a very stringent condition, specially in a situation in which the compact object is modeled as a structure with high density (as neutrons stars, for example). Besides, the isotropic pressure condition becomes unstable by the presence of physical factors such as dissipation, energy density inhomogeneity and shear, this has been recently proven in  \cite{LHP}. Thus, after assuming that the fluid pressure is anisotropic, the two principal stresses (say $P_r$ and $P_\bot$) are unequal and the polytrope equation of state reads
\begin{equation}P_r=K\rho_{0}^{\gamma}=K\rho_0^{1+1/n} .\label{Pol}
\end{equation}
Note that, the introduction of $P_{\perp}$ as a new variable yields an additional degree of freedom and therefore Eq. (\ref{Pol}) is not enough to integrate the  Lane--Emden equation. Thus, in order to decrease the number of degrees of freedom the introduction of an additional equation of state is mandatory. Alternatively,
we can impose certain conditions on the metric variables as the vanishing of the Weyl tensor, implemented in \cite{prnh} to obtain a conformally flat polytrope for anisotropic matter. This condition has its own interest, since it has been seen that highly compact configurations may be obtained with the specific distribution of anisotropy created by such a condition. Other approaches as the Randall-Sundrum model \cite{RS} or $5$-dimensional warped geometries have served as an inspiration for other type of conditions, relating radial derivatives of the metric functions in spherically symmetric spacetimes, that produce self--gravitating spheres embedded in a $5$-dimensional flat space-time (embedding class one). In this regard, it is clear that embedding of four-dimensional spacetimes into higher dimensions is a very useful tool in order to generate astrophysical interior models. Even more, models embedded in five dimensional spacetimes satisfy the so--called Karmarkar or class I condition \cite{karmarkar} (for recent developments see, \cite{Ktello,tello2,tello3,tello4,Nunez,tello5,tello6,maurya,Maurya:2020djz,kk1,kk2,kk3,kk4,kk5,kk6,kk7,kk8,kk9,kk10,kk11,kk12,kk13}, for example) which allows to choose one of the metric functions as the one which generates the total solution. In this work we consider stellar interiors supported by anisotropic fluids fulfilling the polytropic equation of state for the radial pressure and additionally, we implement the class I condition to close the system and construct the corresponding generalized Lane-Emden equation.

The manuscript is organized as follows. In the next section we review the main aspects of anisotropic polytropes and obtain the generalize Lane--Emden equation. Next, we introduce the class I condition and solve the Lane--Emden equation numerically. We conclude this section with a brief discussion about the incidence of the class I condition on the Tolman mass.
The last section is devoted to final remarks and conclusions.

\section{The polytrope for anisotropic fluid}\label{polyaniso}
\subsection{The field equations and conventions}
Let us consider
a static and spherically symmetric distribution of anisotropic matter which metric, in \break Schwarzschild--like coordinates, is parametrized as
\begin{eqnarray}\label{metric}
ds^{2}=e^{\nu}dt^{2}-e^{\lambda}dr^{2}-r^{2}(
d\theta^{2}+\sin^{2}d\phi^{2}),
\end{eqnarray}
where $\nu$ and $\lambda$ are functions of $r$. The metric (\ref{metric}) has to satisfy Einstein field equations \footnote{We are assuming natural units $G=c=1$}
\begin{eqnarray}\label{EE}
G^\mu_\nu = - 8 \pi T^\mu_\nu.
\end{eqnarray}
The matter content of the system is described by the energy--momentum tensor 
\begin{eqnarray}\label{tmunu}
T_{\mu\nu}=(\rho+P_{\perp})u_{\mu}u_{\nu}-P_{\perp}g_{\mu\nu}+(P_{r}-P_{\perp})s_{\mu}s_{\nu},
\end{eqnarray}
where $\rho$ is the total energy density,
\begin{eqnarray}\label{v1}
u^{\mu}=(e^{-\nu/2},0,0,0),
\end{eqnarray}
is the four velocity of the fluid and $s^{\mu}$ is defined as
\begin{eqnarray}\label{v2}
s^{\mu}=(0,e^{-\lambda},0,0),
\end{eqnarray}
with the properties $s^{\mu}u_{\mu}=0$, $s^{\mu}s_{\mu}=-1$. 

Replacing (\ref{metric}), (\ref{tmunu}), (\ref{v1}) and (\ref{v2}) in (\ref{EE}), we have
\begin{eqnarray}
T^0_0=\rho&=&-\frac{1}{8\pi}\bigg[-\frac{1}{r^{2}}+e^{-\lambda}\left(\frac{1}{r^{2}}-\frac{\lambda'}{r}\right) \bigg],\label{ee1}\\
- T^1_1 = P_{r}&=&-\frac{1}{8\pi}\bigg[\frac{1}{r^{2}}-e^{-\lambda}\left(
\frac{1}{r^{2}}+\frac{\nu'}{r}\right)\bigg],\label{ee2}
\end{eqnarray}
\begin{equation}
-T^2_2= P_{\perp}=\frac{1}{8\pi}\bigg[ \frac{e^{-\lambda}}{4}
\left(2\nu'' +\nu'^{2}-\lambda'\nu'+2\frac{\nu'-\lambda'}{r}
\right)\bigg]\label{ee3},
\end{equation}
where primes denote derivative with respect to $r$. Furthermore, we shall consider that
the fluid distribution is surrounded by the Schwarzschild vacuum, namely
\begin{eqnarray}
ds^{2}&=&\left(1-\frac{2M}{r}\right)dt^{2}-\left(1-\frac{2M}{r}\right)^{-1}dr^{2}\nonumber\\
&&-r^{2}(d\theta^{2}+\sin^{2}d\phi^{2}),
\end{eqnarray}
where $M$ represents the total energy of the system.
In order to match the two metrics smoothly on the boundary surface $r=r_{\Sigma}=\rm constant$, we require continuity of the first and second fundamental forms across that surface. As a result of this matching we obtain the well known result,
\begin{eqnarray}
e^{\nu_{\Sigma}}&=&1-\frac{2M}{r_{\Sigma}},\label{nursig}\\
e^{-\lambda_{\Sigma}}&=&1-\frac{2M}{r_{\Sigma}}\label{lamrsig}\\
P_{r_{\Sigma}}&=&0,\label{prr}
\end{eqnarray}
where the subscript $\Sigma$ indicates that the quantity is evaluated at the boundary surface.
From the radial component of the conservation law,
\begin{eqnarray}\label{Dtmunu}
\nabla_{\mu}T^{\mu\nu}=0,
\end{eqnarray}
one obtains  the generalized Tolman--Oppenhei-\break mer--Volkoff equation for anisotropic matter  which reads,
\begin{eqnarray}\label{TOV}
P_{r}'=-\frac{\nu'}{2}(\rho +P_{r})+\frac{2}{r}(P_{\perp}-P_{r}).
\end{eqnarray}
Alternatively, using
\begin{eqnarray}
\nu'=2\frac{m+4\pi P_{r}r^{3}}{r(r-2m)},
\end{eqnarray}
where the mass function $m$ is as usually   defined as
\begin{eqnarray}\label{m}
e^{-\lambda}=1-2m/r ,
\end{eqnarray}
we may rewrite Eq. (\ref{TOV}) in the form
\begin{eqnarray}\label{TOV2}
P_{r}'=-\frac{m+4\pi r^{3} P_{r}}{r(r-2m)}(\rho+P_{r})+\frac{2}{r}\Delta, \label{TOVb}
\end{eqnarray}
where
\begin{eqnarray}\label{anisotropy}
\Delta=P_{\perp}-P_{r},
\end{eqnarray}
measures  the anisotropy of the system.

For the physical variables appearing in (\ref{TOVb}) the following boundary conditions apply
\begin{eqnarray}
 m(0)=0,\qquad m(\Sigma)=M, \qquad P_{r}(r_{\Sigma})=0.
\end{eqnarray}
As already mentioned in the introduction, in order to integrate  equation (\ref{TOVb}), we shall need additional information. In this work  we use the polytropic equation of state for the radial pressure and the class I condition in order to close the system. In the next section we will implement the polytropic equation of state to construct the general Lane--Emden equation.

\subsection{Relativistic polytrope for anisotropic fluids}
When considering the polytropic equation of state within the context of general relativity, two different possibi-\break lities arise, leading to the same equation in the Newtonian limit \cite{herrera2013}. The first one, preserves the original polytropic equation of state (\ref{Pola}) and the second case allows another (natural) possibility that consists in assuming that the relativistic polytrope is defined by,
\begin{eqnarray}
P_{r}=K\rho^{\gamma}=
K\rho^{1+\frac{1}{n}} \label{pr1b}.
\end{eqnarray}
In this case the baryonic density $\rho_0$ is replaced by the total energy density $\rho$ in the polytropic equation of state. The general treatment is very similar for both cases and therefore, for simplicity, we shall restrict here to the case described by (\ref{pr1b}). It can be shown that the relationship between the two densities is given by \cite{herrera2013},
\begin{eqnarray}
\rho= \frac{\rho_0}{\left(1 - K \rho_0^{1/n}\right)^n}
\label{Rho}.
\end{eqnarray}

As it is well known from the general theory of polytropes, there is a bifurcation at the value $\gamma=1$. Thus, the cases $\gamma=1$ and $\gamma\neq1$ have to be considered separately.

Let us first consider the case $\gamma\neq1$. Now, defining the  variable $\psi$ by
\begin{eqnarray}\label{rho1}
\rho = \rho_c \psi^n,
\end{eqnarray}
where   $\rho_{c}$  denotes the energy density at the center  (from now on the subscript $c$ indicates that the variable is evaluated at the center), we may rewrite (\ref{pr1b}) as
\begin{eqnarray}\label{Pr}
 P_{r} = K \rho^{\gamma} = K \rho_c \psi^{n+1} = P_{rc} \psi^{n+1},
\end{eqnarray}
with $P_{rc}=K\rho_{c}^{\gamma_{r}}$. Now, replacing (\ref{pr1b}) and (\ref{rho1}) in (\ref{TOV2}) we obtain
\begin{eqnarray}\label{lmeq}
(n+1)  \psi' &=& - \left(\frac{m + 4 \pi P_{rc} \psi^{n+1}  r^3}{r(r -2m)} \right) \frac{1}{\alpha} \left( 1 + \alpha \ \psi^{n} \right)\nonumber\\
&&\hspace{4.2cm}+ \frac{2 \Delta}{r P_{rc}},
\end{eqnarray}
where $\alpha=P_{rc}/\rho_{c}$.
Let us now introduce the following dimensionless variables
\begin{eqnarray}
     r &=&\frac{\xi}{A}, \quad A^2 = \frac{4 \pi \rho_c}{\alpha(n+1)}, \label{eq:var1} \\
    \psi^n &=& \frac{\rho}{\rho_c}, \quad \eta(\xi) = \frac{m(r)\ A^3}{4 \pi \rho_c},  \label{eq:var2}
\end{eqnarray}
from where (\ref{lmeq}) reads
\begin{eqnarray}\label{lemd}
&&\xi^2 \frac{d\psi}{d\xi} \left[ \frac{1 - 2 \alpha (n+1) \frac{\eta}{\xi}}{1 + \alpha \ \psi } \right]+ \eta + \alpha \xi^3 \psi^{n+1}\nonumber\\
&& - \frac{2 \Delta \ \xi}{P_{rc} \psi^n (n+1)} \left[ \frac{1 - 2 \alpha (n+1) \frac{\eta}{\xi}}{1 + \alpha \ \psi } \right] = 0,
\end{eqnarray}
which corresponds to the generalized Lane--Emden equation for this case. 

Let us now consider the isothermal case: $n=\pm \infty$, $\gamma=1$. In this case, the definition of the variable $\psi$ becomes,
\begin{eqnarray}\label{eq:rho-iso}
    \rho &=& \rho_c e^{-\psi},
\end{eqnarray}
and we can write equation (\ref{pr1b}) as
\begin{eqnarray}
    P_{r} &=& K \rho = K \rho_c e^{-\psi}=P_{rc}e^{-\psi}.   \label{eq:polytrope-iso}
\end{eqnarray}
Replacing (\ref{eq:rho-iso}) and (\ref{eq:polytrope-iso}) in TOV equation (\ref{TOV2}), we obtain
\begin{eqnarray}\label{TOV:iso}
    e^{-\psi} \psi' = \left(\frac{m + 4 \pi P_{rc} e^{-\psi} r^3}{r(r -2m)} \right) \frac{1}{\alpha} \left( e^{-\psi} + \alpha \ e^{-\psi} \right) - \frac{2 \Delta}{r P_{rc}}.\nonumber\\
\end{eqnarray}
Introducing dimensionless variables,
\begin{eqnarray} 
    \alpha = \frac{P_{rc}}{\rho_c}, \quad r =\frac{\xi}{A}, \quad A^2 = \frac{4 \pi \rho_c}{\alpha}, \label{eq:var1-iso} \\
    e^{-\psi} = \frac{\rho}{\rho_c}, \quad \eta(\xi) = \frac{m(r)\ A^3}{4 \pi \rho_c},  \label{eq:var2-iso}
\end{eqnarray}
Eq. (\ref{TOV:iso}) becomes
\begin{eqnarray}\label{TOV:adm}
    \xi^2 \psi'\left(\frac{1 -2 \frac{\alpha}{\xi} \eta}{ 1 + \alpha}\right) &-& \eta - \alpha e^{-\psi} \xi^3\nonumber\\  
    &+& \frac{2 \Delta  e^{\psi}\xi  }{P_{rc}}\left(\frac{1 -2 \frac{\alpha}{\xi} \eta}{ 1 + \alpha}\right) = 0,\nonumber\\\label{eq:lande-iso}
\end{eqnarray}
which corresponds to the Lane-Emden equation for the isothermal case.

\section{The class I condition}\label{modeling}
Embedding of four-dimensional spacetimes into higher dimensions is an invaluable tool in generating both cosmological and astrophysical models \cite{maurya}.
As it is well known, the Karmarkar condition is necessary and sufficient to ensure
class one solutions \cite{karmarkar} which 
 for spherically symmetric space--times reads ($R_{\theta\phi\theta\phi}\neq0$) \cite{karmarkar}
\begin{equation}\label{eq15}
R_{r t r t}R_{\theta\phi \theta\phi}=R_{r\theta r \theta}R_{t \phi t \phi}+R_{r\theta t \theta}R_{r\phi t\phi},    
\end{equation}
leading to
\begin{equation}\label{eq16}
2\frac{\nu^{\prime\prime}}{\nu^{\prime}}+\nu^{\prime}=\frac{\lambda^{\prime}e^{\lambda}}{e^{\lambda}-1},
\end{equation}
with $e^{\lambda}\neq 1$.
Now,  using (\ref{rho1}), (\ref{Pr}) and (\ref{eq16}) in
(\ref{anisotropy}) we obtain
\begin{eqnarray}\label{anis:dimen}
 \Delta =   \frac{ \left( 4 \pi P_{rc} \psi^{n+1} r^3 - m \right)\left(r m' - 3 m \right)  }{16 \pi m\ r^3 },
\end{eqnarray}
from where, after using the set of dimensionless variables (\ref{eq:var1}) and (\ref{eq:var2}), we arrive at
\begin{eqnarray}\label{anisC1}
 \Delta = \rho_c \frac{\left( \alpha \psi^{n+1} \xi^3 -  \eta \right)\left(\xi \eta' - 3 \eta   \right)}{ 4 \xi^3 \eta}.
\end{eqnarray}
Finally, replacing (\ref{anisC1}) in (\ref{lemd}), the class I generalized Lane-Emden equation for $\gamma \ne 1$ reads
\begin{eqnarray} \label{eq:lande}
&&\frac{ \left( \alpha \psi^{n+1} \xi^3 -  \eta \right)\left(3 \eta - \xi \eta'  \right) }{2 \alpha(n+1)\  \xi^2 \psi^n \eta } \left[ \frac{1 - 2 \alpha (n+1) \frac{\eta}{\xi}}{1 + \alpha \ \psi } \right]\nonumber\\
&&+\xi^2 \psi' \left[ \frac{1 - 2 \alpha (n+1) \frac{\eta}{\xi}}{1 + \alpha \ \psi } \right] + \eta + \alpha \xi^3 \psi^{n+1}= 0. \nonumber\\
\end{eqnarray}
We can repeat the same procedure for the $\gamma=1$ (isothermal) case. Again, after using (\ref{eq:rho-iso}), (\ref{eq:polytrope-iso}) and (\ref{eq16}) in (\ref{anisotropy}) we obtain,
\begin{eqnarray}\label{delta:iso}
 \Delta = \frac{ \left( 4 \pi P_{rc} e^{-\psi} r^3 - m \right)\left(r m' - 3 m \right)  }{16 \pi m\ r^3 },
\end{eqnarray}
from where, by means of the definitions (\ref{eq:var1-iso}) and (\ref{eq:var2-iso}) we have
\begin{eqnarray}\label{delta:iso:final}
 \Delta = \rho_c \frac{\left( \alpha e^{-\psi} \xi^3 -  \eta \right)\left(\xi \eta' - 3 \eta   \right)}{ 4 \xi^3 \eta}. 
 \label{eq:ani_norm-iso}
\end{eqnarray}
Finally, introducing (\ref{delta:iso:final}) in (\ref{TOV:adm}) we arrive to the class I generalized Lane-Emden equation for $\gamma=1$ (isothermal case),
\begin{eqnarray}
 &&\xi^2 \frac{d\psi}{d\xi}\left(\frac{1 -2 \frac{\alpha}{\xi} \eta}{ 1 + \alpha}\right) - \eta - \alpha e^{-\psi} \xi^3  \nonumber\\
 &+& e^{\psi}  \frac{\left( \alpha e^{-\psi} \xi^3 -  \eta \right)\left(\xi \eta' - 3 \eta   \right)}{ 2 \alpha \xi^2 \eta}  \left(\frac{1 -2 \frac{\alpha}{\xi} \eta}{ 1 + \alpha}\right) = 0.\nonumber\\
 \label{eq:lande-iso-final}
\end{eqnarray}

To complement the discussion, we proceed to calculate the Tolman mass (for the $\gamma\ne 1$ case) \cite{LHTolman}, defined by
\begin{equation}
    m_{T}=\frac{1}{2}e^{(\nu-\lambda)/2} \nu^\prime  r^{2},
    \label{eq:mtw}
\end{equation}
or, alternatively \cite{herrera97}
\begin{equation}
       m_{T}=e^{(\nu+\lambda)/2} \left(m + 4 \pi r^3 P_r\right).\\
       \label{eq: mtw-m}
    \end{equation}
The functions $\lambda$, $m$ and $P_r$ in the above expression, are obtained directly by integration of  equation (\ref{eq:lande}) for the $\gamma\ne 1$ case. So, we only need an expression for $\nu$, which can be obtained following the systematic procedure described in \cite{herrera2013,doble_P}. Using the corresponding expressions (\ref{rho1}) and (\ref{Pr}), and the variables defined in (\ref{eq:var1}) (\ref{eq:var2}), we have
    \begin{eqnarray}
         m_{T} &=&e^{(\nu+\lambda)/2}\left(\alpha (n+1)\eta +\alpha^2 (n+1) \psi^{n+1} \xi^3\right)\frac{1}{A}.\nonumber\\\label{eq:mtw_0}
    \end{eqnarray}
Now, in order to obtain the metric variable $\nu$, we proceed as follows. First, the TOV
equation (\ref{TOV2}) can be written as
  \begin{eqnarray}
    2\alpha (n+1) d\psi &=& -\left(\alpha \psi + 1 \right) d\nu + 4 \frac{ \Delta}{r \rho_c \psi^n}dr ,
\end{eqnarray}
from where
\begin{eqnarray}
    &&2\alpha (n+1) \int_{\psi(r)}^{\psi(r_{\Sigma})} \frac{d\psi}{\alpha \psi + 1} = \nonumber\\
    &-&\int_{\nu(r)}^{\nu(r_{\Sigma})} d\nu + \frac{4}{\rho_c} \int_{r}^{r_{\Sigma}} \frac{ \Delta}{\psi^n \left(\alpha \psi + 1 \right) r}dr. \label{eq:diferencial}
\end{eqnarray}
Next, defining $G(r)$ as,
\begin{equation*}
    G(r)\equiv \int_{r}^{r_{\Sigma}} \frac{ \Delta}{\psi^n \left(\alpha \psi + 1 \right) r}dr ,
\end{equation*}
the integration of (\ref{eq:diferencial}) produces
\begin{eqnarray}
    -2\alpha (n+1) \log{(\alpha \psi+1)}=\nu(r)-\nu(r_{\Sigma})+\frac{4}{\rho_c}G(r),
\end{eqnarray}
from where we obtain
\begin{eqnarray}
    e^{\nu(r)}&=&\frac{e^{\nu(r_{\Sigma})}}{(\alpha \psi+1)^{2\alpha(1+n)} e^{\frac{4}{\rho_c}G(r)}}. \label{eq:nu}
\end{eqnarray}
Using (\ref{m}) and (\ref{eq:nu}) in (\ref{eq:mtw_0}) we arrive at
\begin{eqnarray}
          m_{T}&&=\left(\frac{a_{\Sigma}}{a}\right)^{1/2}(1+\alpha\psi)^{-\alpha(n+1)}e^{-\frac{2}{\rho_c}G(r)} \times \nonumber\\
          &&\left[\alpha (n+1)\eta +\alpha^2 (n+1) \psi^{n+1} \xi^3\right]\left(\frac{\alpha (n+1)}{4\pi \rho_{c}}\right)^{1/2}\nonumber\\\label{eq:final2-tw}
\end{eqnarray}
where
    \begin{eqnarray}
        a=1-2\alpha(n+1)\frac{\eta(\xi)}{\xi}.
    \end{eqnarray}
For the numerical calculations it is convenient to change to the following dimensionless variables
\begin{eqnarray}
    x &=& \frac{r}{r_\Sigma} = \frac{\xi}{r_\Sigma A}=\frac{\xi}{\xi_\Sigma},\\ 
        y &=& \frac{M}{r_\Sigma} = \alpha (n+1) \frac{\eta_\Sigma}{\xi_\Sigma} \\
        \eta_T &=& \frac{m_T}{4 \pi \rho_c \alpha^3},      
 \end{eqnarray}
 in terms of which equation \eqref{eq:final2-tw} reads,
   \begin{eqnarray}\label{MTolman}
       \eta_{T}=&&\left(\frac{1 - 2 y}{1 - 2 \alpha (n+1) \frac{\eta}{x \xi_\Sigma}}\right)^{1/2}\times\nonumber \\   
       &&\quad\frac{\left[\alpha (n+1)\eta +\alpha^2 (n+1) \psi^{n+1} x^3 \xi_\Sigma^3 \right]}{(1+\alpha\psi)^{\alpha(n+1)}e^{\frac{2}{\rho_c}G(x)}}\times\nonumber\\
       &&\qquad\qquad\qquad\qquad\left(\frac{1}{4\pi \rho_{c} \alpha } \right)^{3/2}  (n+1)^{\frac{1}{2}}.\nonumber\\
   \end{eqnarray}
 where $G(x)$ becomes,
 \begin{equation}\label{G}
    G(x) = \int_{x}^1 \frac{\Delta(x)}{\psi^n (\alpha \psi +1 )x \xi_\Sigma} dx.
\end{equation}
This procedure, where the integration of the TOV equation is carried out, allows us to find the metric function $\nu$ (using the appropriate boundary conditions) and finally Tolman's mass, (\ref{MTolman}) and (\ref{G}), is totally equivalent to the one developed in \cite{doble_P} for the different cases exposed. 

Now, the above results will be integrated numerically,
with the corresponding boundary conditions, in order to obtain the solution for each case considered.\\

\begin{figure}[ht!]
    \centering
    \includegraphics[width = 0.42\textwidth]{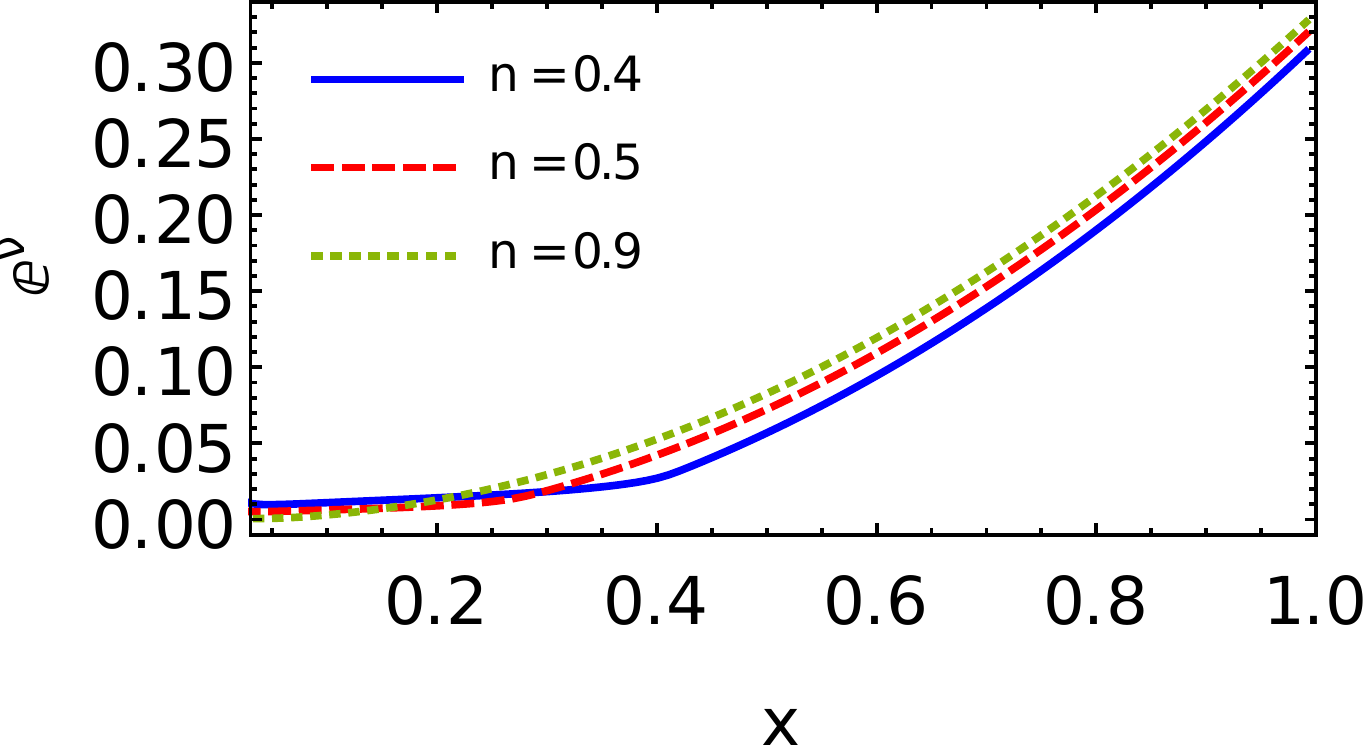}
    \caption{$e^{\nu}$ as function of x for $\alpha = 0.2$ and different values of n.}
    \label{fig:expnu}
\end{figure}

\begin{figure}[ht!]
    \centering
    \includegraphics[width = 0.4\textwidth]{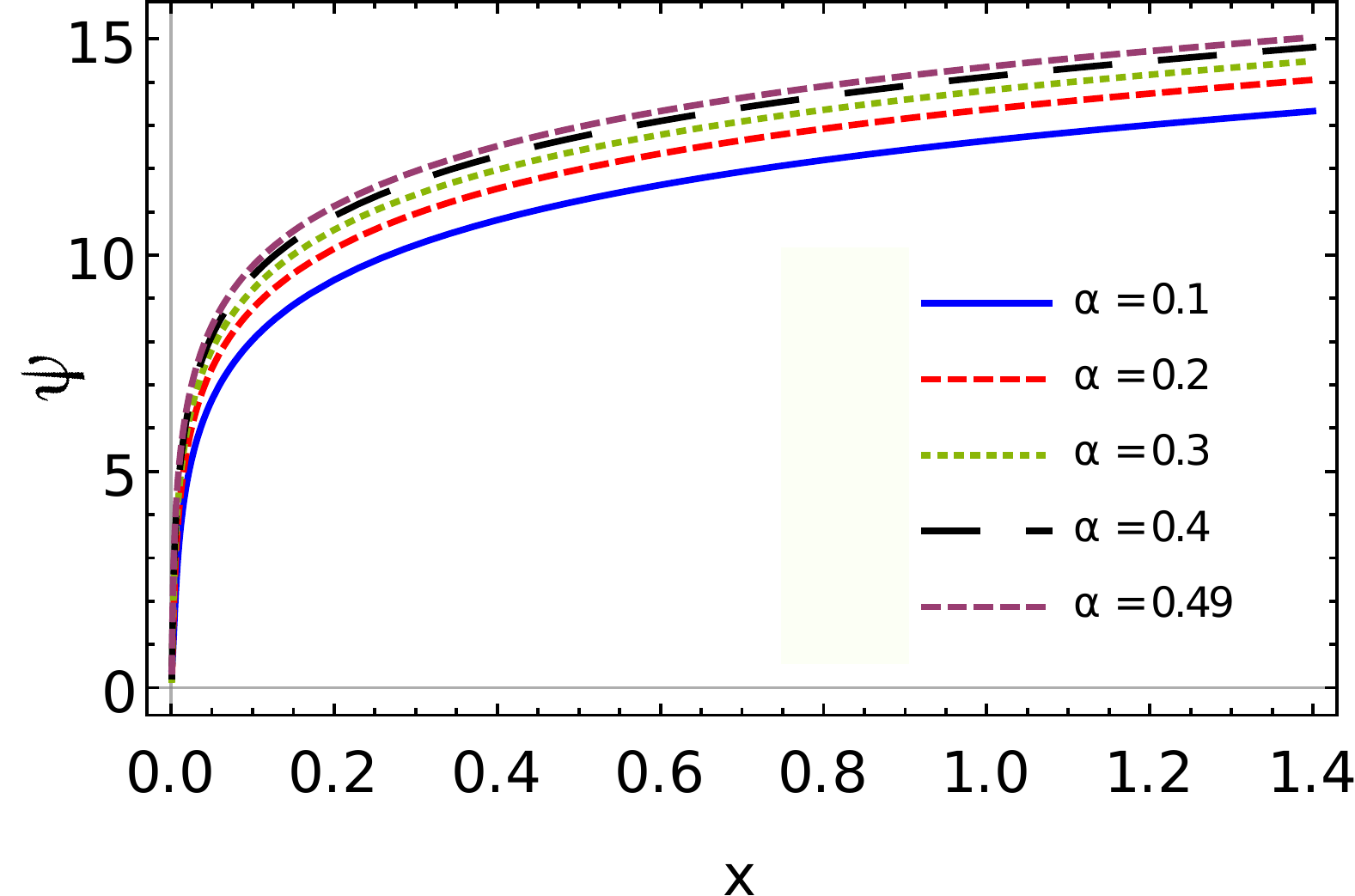}
    \caption{$\psi$ as function of x ( $\gamma = 1$ isothermal case) for different values of $\alpha$}
    \label{fig:iso}
\end{figure}

\begin{figure}[ht!]
    \centering
    \includegraphics[width = 0.4\textwidth]{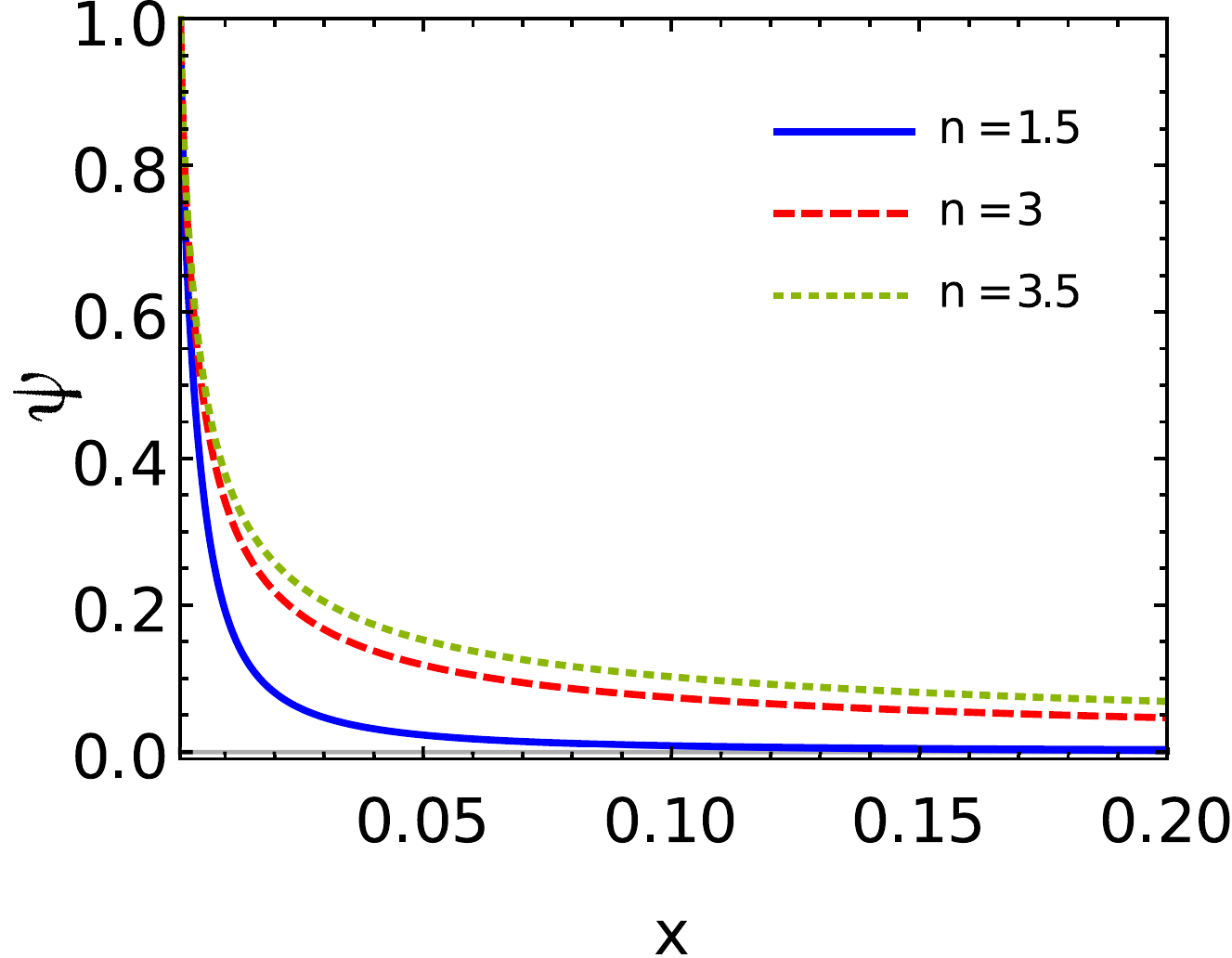}
    \caption{$\psi$ as function of x for $\alpha = 0.1$ and different values of n.}
    \label{fig:psi}
\end{figure}

\begin{figure}[ht!]
    \centering
    \includegraphics[width = 0.4\textwidth]{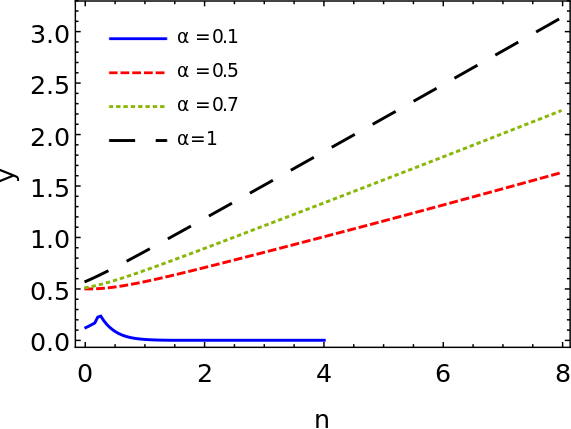}
    \caption{$y$ as function of n for different values of $\alpha$. }
    \label{fig:surf}
\end{figure}

Figure \ref{fig:expnu} shows the behaviour of $e^{\nu}$ as a function of $x$ for different values of the parameters involved. Note that, the function is monotonously increasing as required. In figures \ref{fig:iso} and \ref{fig:psi} it is shown the integration of Eqs. (\ref{eq:lande-iso-final}) (isothermal case) and (\ref{eq:lande}), respectively, for the values of the parameters indicated in the figure legend. Note that for the isothermal case ($\gamma=1$)
the configurations are unbounded  and as a consequence, it is meaningless to define a surface potential or the total mass. However, for the $\gamma\ne 1$ case $\psi$ is monotonously decreasing as expected, and the radial pressure $P_r$ vanishes at the surface as required by the continuity of the second fundamental form. 
The parameter $y$ (``the surface potential'') which measures the degree of compactness is plotted in Fig. \ref{fig:surf}
 as function of the polytropic index $n$ for different values of $\alpha$. Note that the plots have discontinuities in some point $n_d$, where we observed that for $n < n_{d}$, the solution is well--behaved, and for $n \ge n_{d}$, the numerical solution starts to have either imaginary components or strange oscillatory behavior.
 
\begin{figure}[ht!]
    \centering
    \includegraphics[width = 0.4\textwidth]{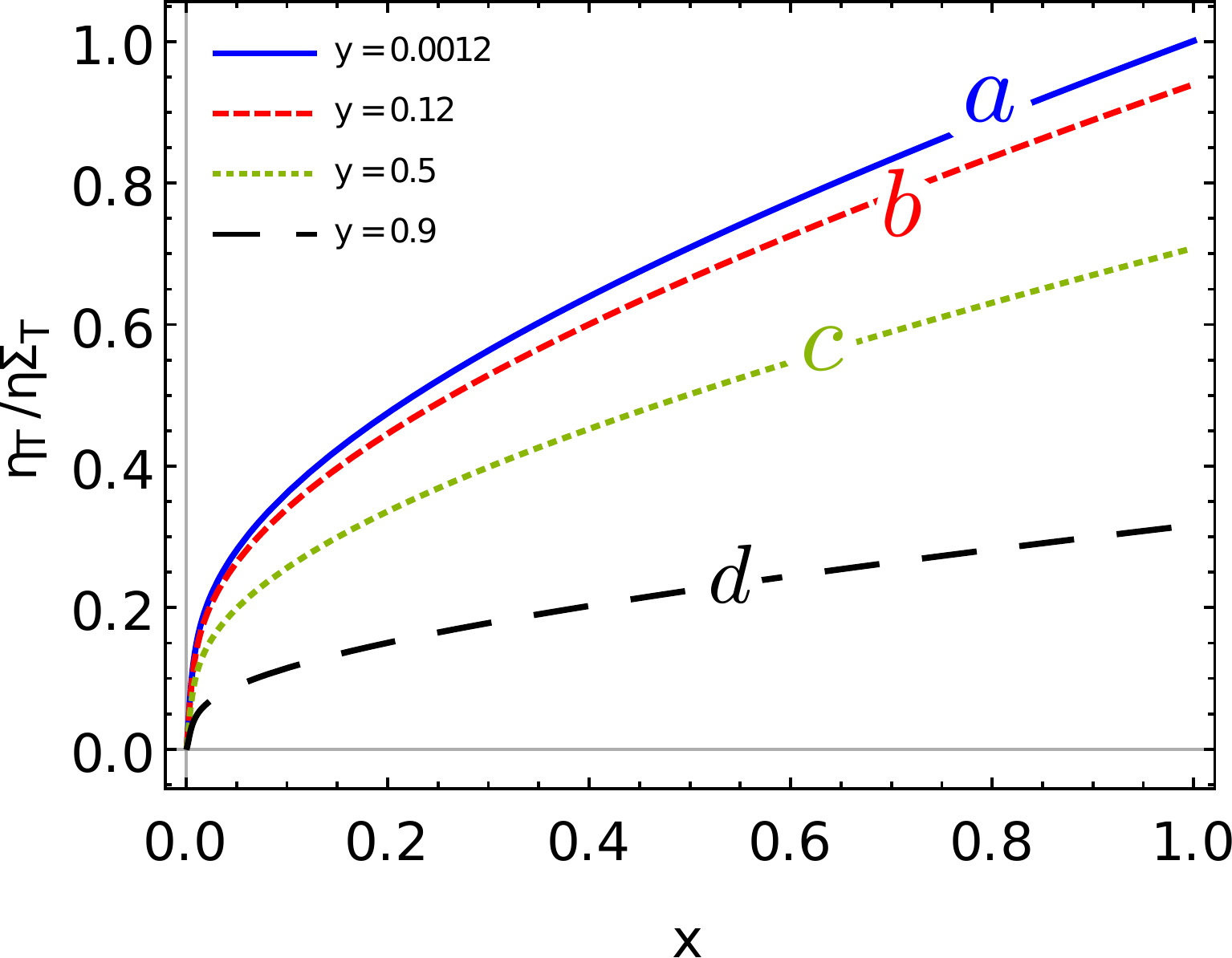}
    \caption{Normalized Tolman-Whittaker Mass $\eta_{T}/\eta_{\Sigma_{T}}$ as function of $x$ for $\alpha = 0.1$, $n= 3$ and different values of $y$. Values of y are read off figure \ref{fig:surf}.}
    \label{fig:mT1}
\end{figure}

\begin{figure}[ht!]
    \centering
    \includegraphics[width = 0.4\textwidth]{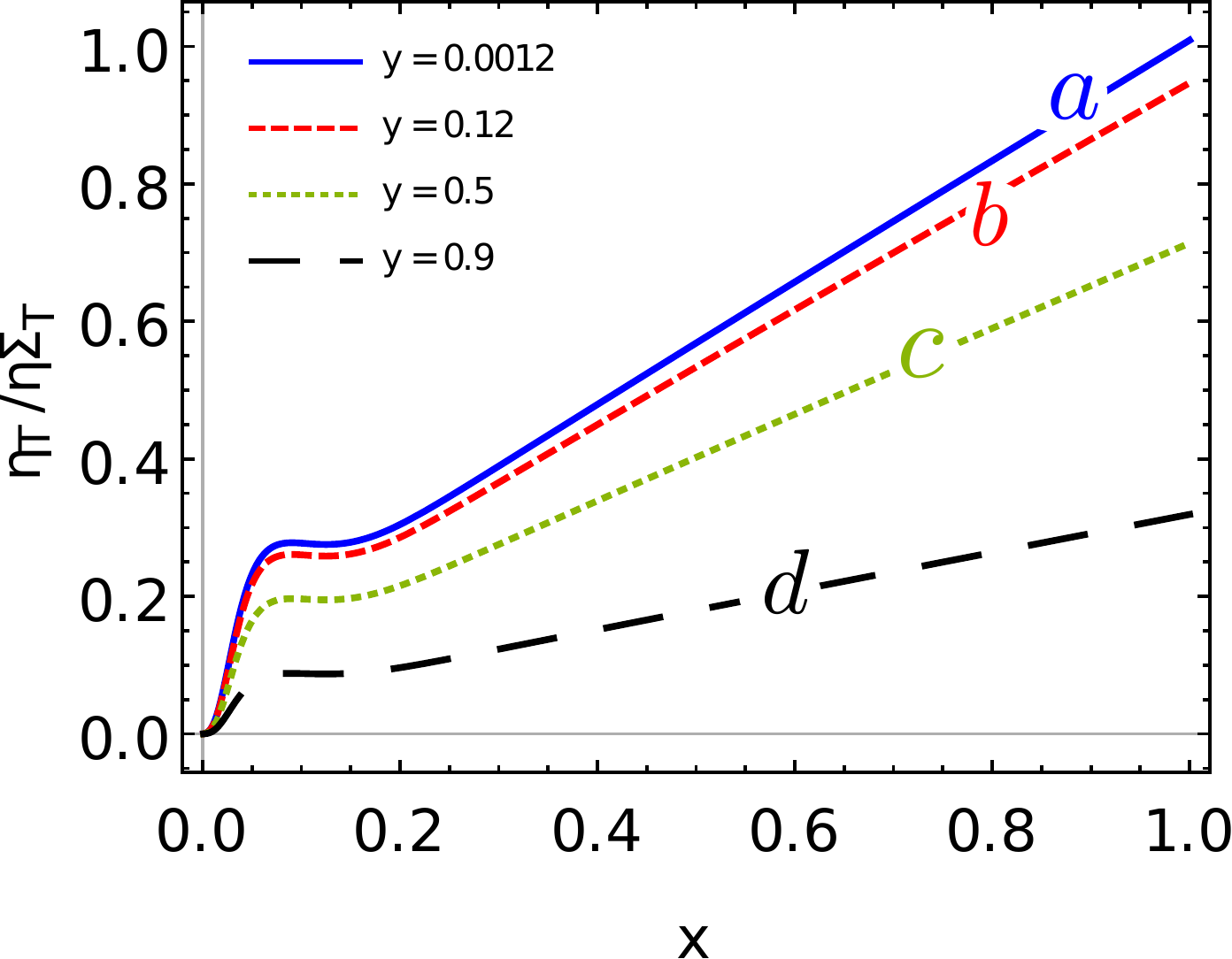}
    \caption{Normalized Tolman-Whittaker Mass $\eta_{T}/\eta_{\Sigma_{T}}$ as function of $x$ for $\alpha = 0.1$, $n= 1$ and different values of $y$. Values of y are read off figure \ref{fig:surf}.}
    \label{fig:mT2}
\end{figure}

\begin{figure}[ht!]
    \centering
    \includegraphics[width = 0.4\textwidth]{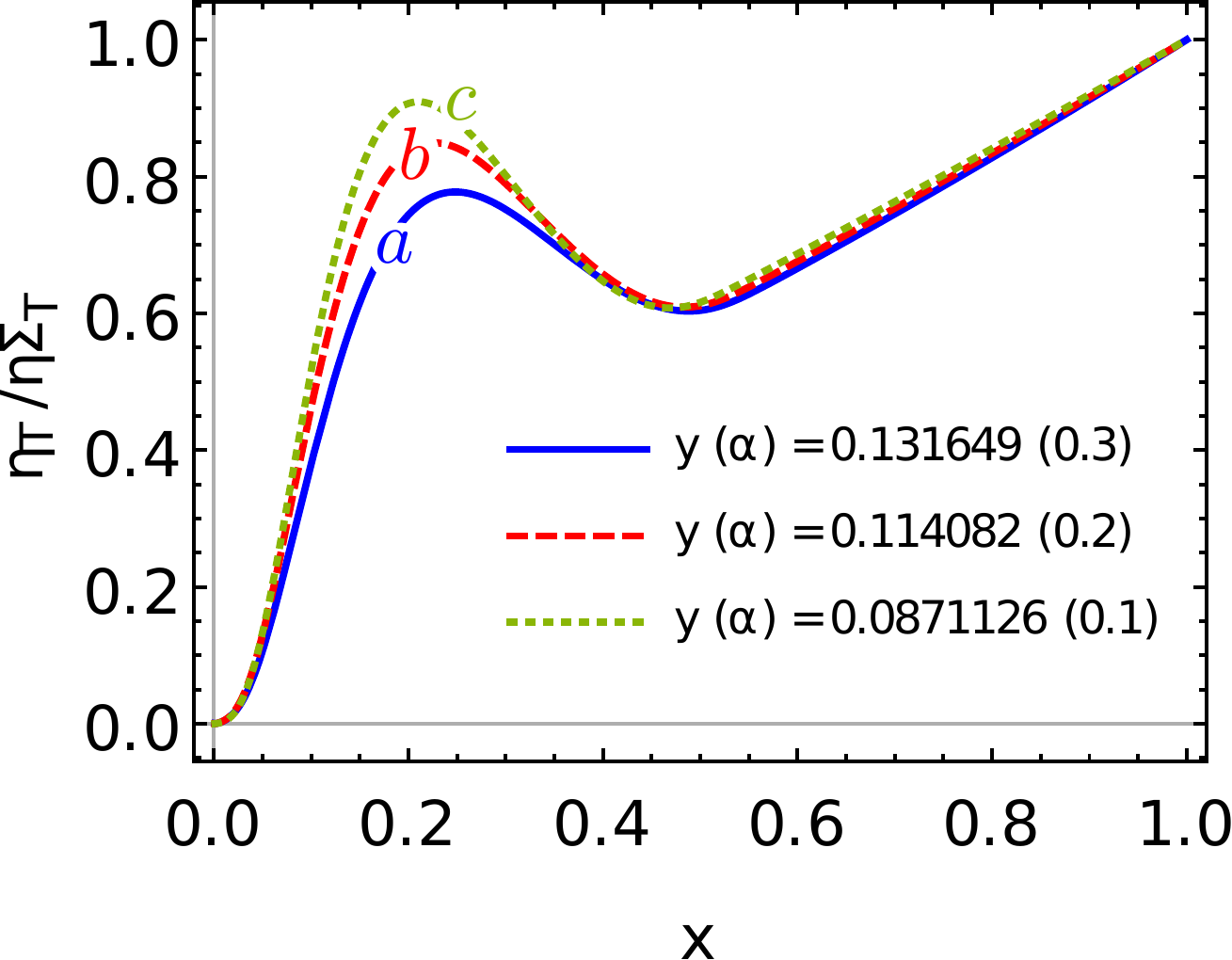}
    \caption{Normalized Tolman-Whittaker Mass  $\eta_{T}/\eta_{\Sigma_{T}}$ as function of $x$ for $n = 0.5$ and different values of $y$ $(\alpha)$. (Blue) $y$ $(\alpha)$ = 0.131649 (0.3). (Red) $y$ $(\alpha)$ = 0.114082 (0.2). (Green) $y$ $(\alpha)$ = 0.0871126 (0.1). Values of y are read off figure \ref{fig:surf}.}
    \label{fig:mT}
\end{figure}

Figures \ref{fig:mT1} and \ref{fig:mT2} display the Tolman mass (normalized by the total mass), for the case $\gamma \ne 1$, as function of $x$ for the selection of values of the parameters indicated in the legend. The behavior of the curves is qualitatively the same for a wide range of values of the parameters. We observe that as we move from the less compact configuration (curve a) to the more compact one (curve d), the Tolman mass tends to concentrate on the outer regions of the sphere. Therefore, it may be inferred from the figures \ref{fig:mT1} and \ref{fig:mT2} that more stable configurations correspond to smaller values of $y$ since they are associated to a sharper reduction of the Tolman mass in the inner regions, as usually happens if the anisotropy is fixed by the imposed condition. This is the usual strategy adopted by the fluid distribution to keep the equilibrium; it tries to concentrate the Tolman mass in the outer regions. This behavior was already observed for different families of anisotropic polytropes discussed in \cite{herrera2013}. The same result is obtained from the Fig. \ref{fig:mT} where we plot the Tolman mass as a function of $x$ for the selection of values of the parameters indicated in the legend.

\begin{figure}[ht!]
    \centering
    \includegraphics[width = 0.4\textwidth]{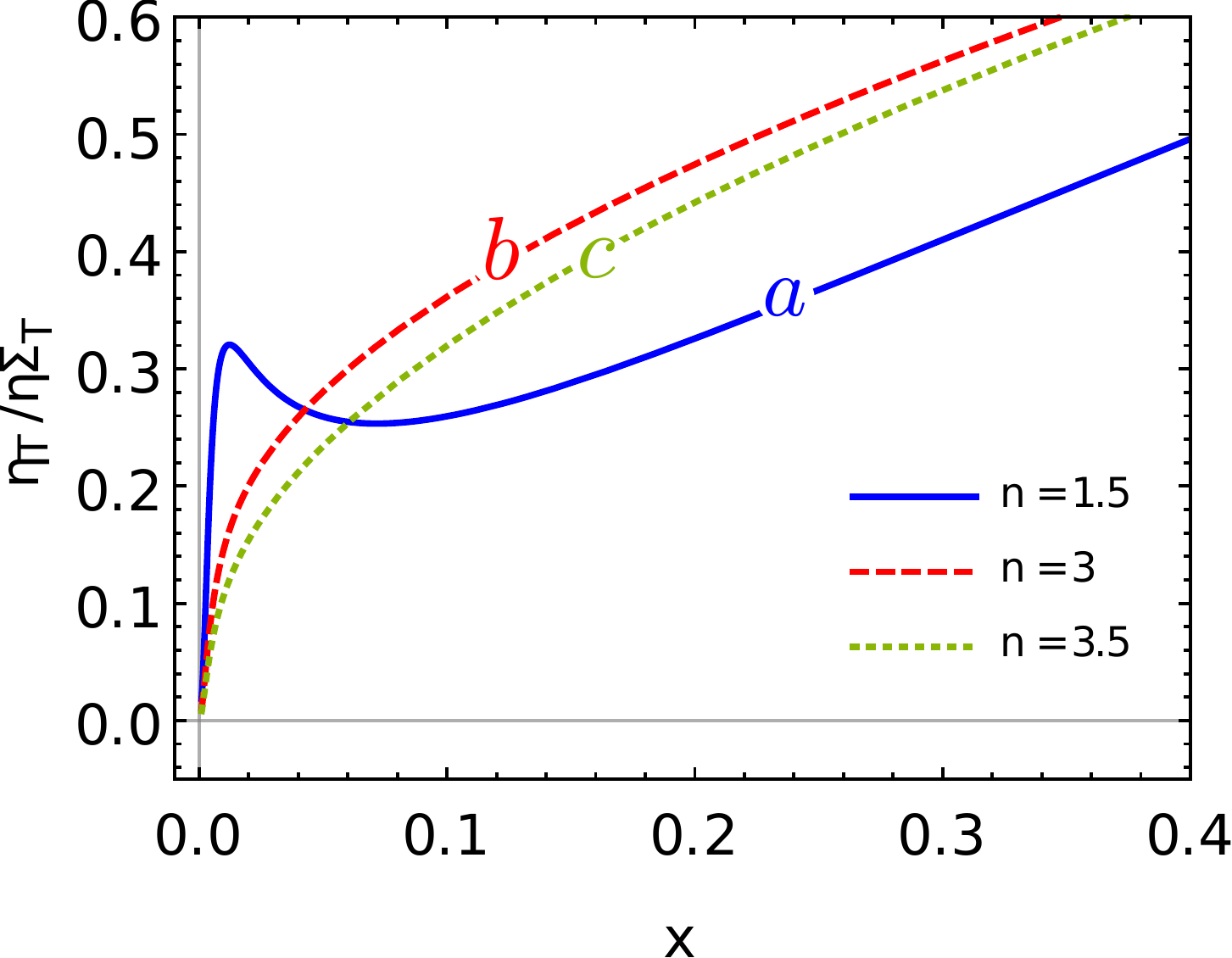}
    \caption{Normalized Tolman-Whittaker Mass  $\eta_{T}/\eta_{\Sigma_{T}}$ as function of $x$ for $\alpha = 0.1$ and different values of $n$. (Blue) $n = 1
1.5$. (Red) $n = 3$. (Green) $n = 3.5$. Values of y are read off figure \ref{fig:surf}.}
    \label{fig:mT4}
\end{figure}

For completeness, we plot the Tolman mass, as function of $x$, for several polytropic indexes and compactness parameters represented by $y$. The result i shown in Fig. \ref{fig:mT4} for the selection of values of the parameters indicated in the legend.

It is worth mentioning that 
all the numerical results show well-behaved solutions for small alpha values and a specific range of values for $n$.

\section{Karmarkar condition and Tolman mass}
In this section we would like to establish some relationships that may be useful to analyze the anisotropy inherited by the Karmarkar condition. In order to relate the influence of the Karmarkar condition with the mass function $m$ and the Tolman mass, $m_{T}$, we will take an instructive path to complete our analysis of the fluid distribution of the compact body. From (\ref{m}) we obtain the usual definition of the mass function,
\begin{equation}
    m = \frac{1}{2} r R^{3}_{232}.
    \label{eq:mRiemann}
\end{equation}
Note that, we can write condition (\ref{eq15}), for our static and spherically symmetric case, in order to define a new variable, $\mathbb{K}$, in the following form,
\begin{equation}
    \mathbb{K} = R_{rtrt} R_{\theta \phi \theta \phi} - R_{r \theta r \theta} R_{\phi t \phi t}, \label{eq:k_karm}
\end{equation}
where the Karmarkar condition is satisfied when $\mathbb{K} = 0$. In this way we can keep track of how this condition influences our anisotropic distribution of fluid of the compact object. Using the components of the Riemann tensor appropriate to the symmetry we may obtain from (\ref{eq:k_karm}) the following expression,
\begin{eqnarray}
    \mathcal{K} &\equiv &  - \frac{\mathbb{K}}{\sin^2 \theta} =   \nonumber\\
    && - e^{(\nu+\lambda)}  \frac{1}{4} e^{-\lambda}  \left( 2\nu^{''} - \lambda' \nu' + \nu'^2 + 2\frac{\nu' - \lambda'}{r} \right)  2 m r\nonumber\\ &+&  \left( e^{(\nu+\lambda)} e^{-\lambda} \frac{\nu' - \lambda'}{2r} \right)   2 m r +\frac{r^2}{4} \lambda' \nu' e^{(\nu-\lambda)} , \label{eq:20}
\end{eqnarray}
which, after using Einstein equations (\ref{ee1}), (\ref{ee2}) and (\ref{ee3}), can be written as
\begin{eqnarray}
    \mathcal{K} &=&   - e^{(\nu+\lambda)} \left[ \left( 8 \pi \left( T^0_0 + T^1_1  -   2 T^2_2 \right) \right) + \frac{4 m}{r^3}  \right]   m r\nonumber\\
    &+&\frac{r^2}{4} \lambda' \nu' e^{(\nu -\lambda)}. \label{eq:22}
\end{eqnarray}
Now, we proceed in two different ways in order to interpret the influence of the Karmarkar condition over our interior solution. First, from (\ref{ee1}) and (\ref{ee2}), we obtain the derivatives of the metric functions $\nu'$ and $\lambda'$ in the form
\begin{eqnarray}
   - \lambda' &=& \frac{2 m}{r^2} e^{\lambda}  - 8 \pi r T^0_0 e^{\lambda},\\ \label{eq:lambda'}
    e^{- \lambda} \nu' &=& \frac{2m}{r^2} - 8 \pi r T^1_1 . \label{eq:nu'}
\end{eqnarray}
Now, replacing these expressions in the last term of (\ref{eq:22}) we obtain 
\begin{eqnarray}\label{k:vs:m}
     \frac{3}{r^2} m^2 &+& 4 \pi r\left(4 T^2_2 - T^1_1 - T^0_0 \right) m \\\nonumber
     &-& 16 \pi^2 r^4 T^0_0 T^1_1 = e^{-(\nu  + \lambda)}\; \mathcal{K},
\end{eqnarray}
which relates the mass function $m$ with the $\mathcal{K}$ condition, defined in (\ref{eq:20}).
In contrast, if we use (\ref{eq:mtw}) and Einstein equation (\ref{ee1}), we can write the last term of (\ref{eq:22}) in terms of the Tolman mass function, $m_{T}$, as
\begin{eqnarray}
    \mathcal{K} &=&   - e^{(\nu+\lambda)} \left[ \left( 8 \pi \left( T^0_0 + T^1_1  -   2 T^2_2 \right) \right) + \frac{4 m}{r^3}  \right]   m r\nonumber\\
    &+& \left[\frac{4\pi T^0_0  \, r^3 - m}{r(r-2m)}\right]  e^{(\nu -\lambda)/2}\, m_{T}. \label{eq:22:mtw}
\end{eqnarray}
Finally, using (\ref{eq: mtw-m}) in the previous equation we have
\begin{eqnarray}
    r \mathcal{K} &=&   - e^{(\nu+\lambda)} \left[ \left( 8 \pi r^2 \left( T^0_0 + T^1_1  -   2 T^2_2 \right) \right) + \frac{4 m}{r}  \right]   m \nonumber\\
    &+& \left[\frac{4\pi r^3 (T^0_0-  T^1_1 )- m_{T}e^{-(\nu +\lambda)/2}}{(r-2m)}\right]  e^{(\nu -\lambda)/2} m_{T},\nonumber\\ \label{eq:22:mtw:m}
\end{eqnarray}
which constitutes a relation between the mass function, the Tolman mass and the $\mathcal{K}$ condition. So, we have been successful in relating the components of the energy--momentum tensor, the mass function and the Tolman mass with the defined $\mathcal{K}$ condition for the symmetry considered in our development. This could allow us to discover the influence over these mass functions by imposing the Karmarkar condition and then be able to find out how this metric (embedding) condition influences the distribution of our compact fluid.

In order to illustrate the effect of the Tolman mass on $\mathcal{K}$ we shall consider a simple example, namely, an anisotropic solution with vanishing radial pressure \cite{Casadio:2019usg},
\begin{eqnarray}
e^{\nu}&=&B^{2}\left(1+\frac{r^{2}}{A^{2}}\label{1}\right)\\
e^{-\lambda}&=&\frac{A^{2}+r^{2}}{A^{2}+3r^{2}}\label{2}\\
\rho&=&\frac{6(A^{2}+r^{2})}{8\pi(A^{2}+3r^{2})^{2}}\label{3}\\
p_{t}&=&\frac{3r^{2}}{8\pi(A^{2}+3r^{2})^{2}}\label{4}\\
p_{r}&=&0\label{5},
\end{eqnarray}
with
\begin{eqnarray}
\frac{A^{2}}{r_{\Sigma}^{2}}&=&\frac{r_{\Sigma}-3M}{M}\label{6}\\
B^{2}&=&1-\frac{3M}{r_{\Sigma}}\label{7},
\end{eqnarray}
which ensure the fulfilling of the matching conditions (\ref{nursig}), (\ref{lamrsig}) and (\ref{prr}). Next, replacing (\ref{1}), (\ref{2}), (\ref{3}), (\ref{4}), (\ref{5}), (\ref{6}) and (\ref{7}) in (\ref{eq:22:mtw:m}) and using (\ref{eq: mtw-m}) to express the mass function in terms of $m_{T}$ we arrive at
\begin{eqnarray}
\mathcal{K}&=&
-\frac{\eta}{\vartheta}  \bigg(
-8 \Xi \eta ^2 \Lambda^2+3 \Xi x^4 y \left(\Lambda^2-24 \eta ^2 y\right)\nonumber\\
&&+24 \Xi \Lambda x^6 y^2-48 \Xi \eta ^2 \Lambda x^2 y+15 \eta  \Theta x^5 y^2\nonumber\\
&&+33 \Xi x^8 y^3+5 \eta  \Theta \Lambda^2 x+16 \eta  \Theta \Lambda x^3 y\bigg),
\end{eqnarray}
where $x=r/_{\Sigma}$, $\eta=m_{T}/_{\Sigma}$,
$y=M/r_{\Sigma}$ and
\begin{eqnarray}
\Xi&=&\sqrt{\frac{\left(x^2-3\right) y+1}{3 \left(x^2-1\right) y+1}}\\
\Theta&=&\sqrt{\left(x^2-3\right) y+1}\\
\Lambda&=&1-3 y\\
\vartheta&=&
\left(3 \left(x^2-1\right) x y+x\right)^2 (\Theta x-2 \Xi \eta )
\end{eqnarray}

\begin{figure}[ht!]
    \centering
    \includegraphics[scale=0.4]{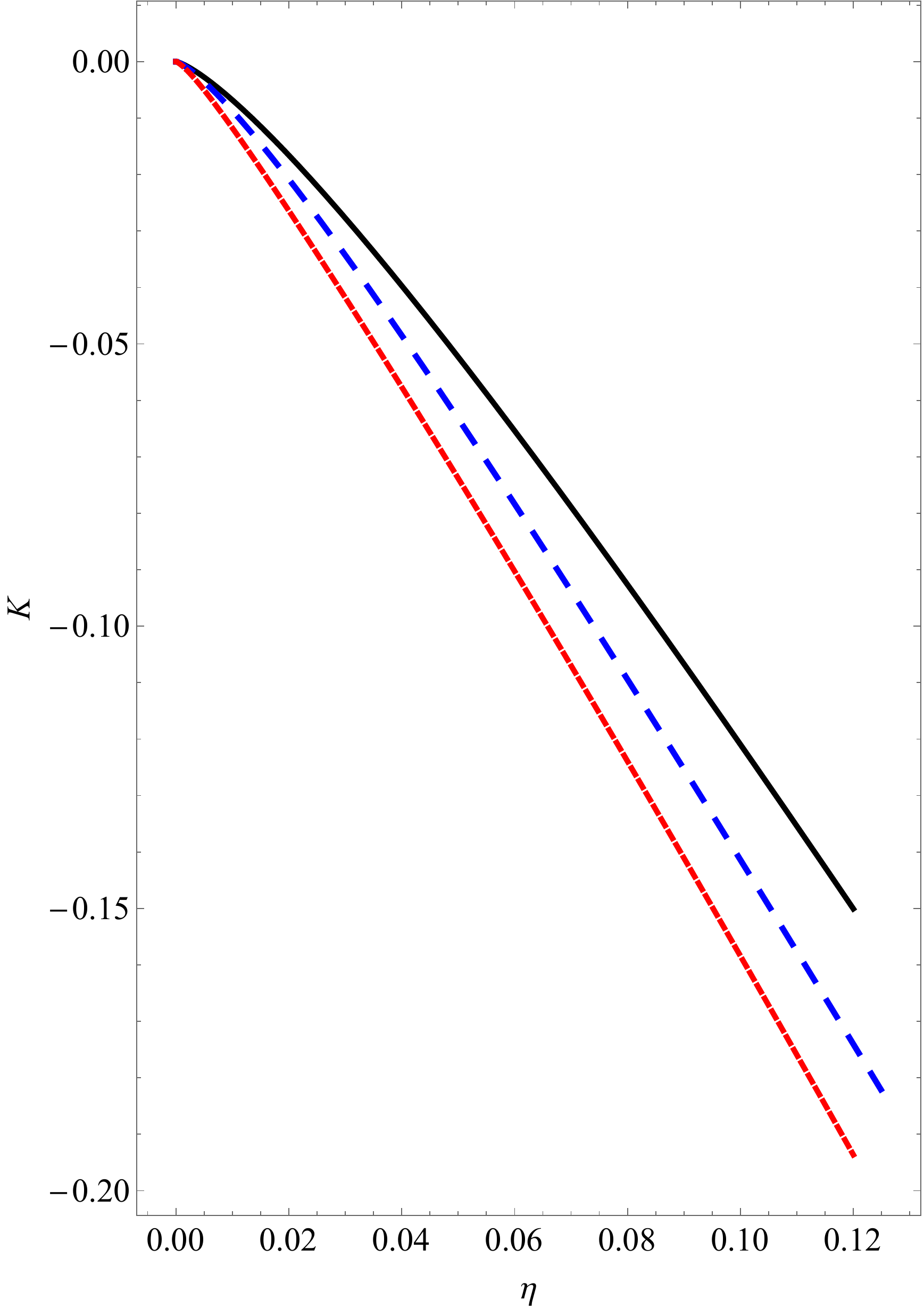}
    \caption{\label{kt}
    Parametric plot of $\mathcal{K}$ as a function of $\eta$ for $y=0.2$ (black line), $y=0.25$ (blue dashed line) and $y=0.3$ (red dashed line)}
    \label{fig:mT41}
\end{figure}

In Fig. \ref{kt}, we show the parametric plot of $\mathcal{K}(x)$ as a function of the normalized Tolman mass, $\eta(x)$ in the interval $x\in(0,1)$ for different values of $y$.
Interestingly, the solution deviates from the $\mathcal{K}=0$ condition as $y$ increases.

\section{Conclusions}

In this work, we have developed a general method to construct locally anisotropic polytropes, and we have applied it to the specific case of a class I solution for sherically symetric interior space-times. As it occurs in the specific case of conformally flat polytrope spheres \cite{prnh}, our models are necessarily anisotropic (principal stresses unequal) and therefore are not continuously linked to isotropic polytropes (just like it ocurres in other reported anisotropic distributions \cite{LH:ponce}). Since the inclusion of pressure anisotropy implies an additional degree of freedom, the integration of the ensuing Lane– Emden equation, in the general case, requires additional information. Usually this information is provided through equations of state or other type of assumptions about the nature of the anisotropy. Here, we have supplied such additional information by assuming that the polytrope has the specific anisotropy derived from de Karmarkar condition. This represents an heuristic assumption whose validity will be confirmed (or not) from its application to specific problems. Then, the geometrical constraint, chosen to work with, produce a specific type of anisotropy leading to a Lane-Emden equation that required a numerical analysis, given the high nonlinearity of the equations.

We could differentiate two possible cases depending on whether $\gamma\ne 1$ or $\gamma = 1$. We have integrated the Lane–Emden equations for these cases, for a very large set of values of the parameters, even though, only a very specific set for each case is exhibited, this because the qualitative behavior of the system does not change much for a wide range of the parameters. The main reason to present such models was not to describe any specific astrophysical scenario, but to illustrate the applications of our approach. However, the protocol here developed could be useful for dealing with several phenomena in which anisotropic polytropes appear, with special emphasis in the physics of compact objects and so on, the obtained models exhibit some interesting features which deserve to be commented. We do not know if distributions that are usually represented as polytropes, and that also constitute very compact objects, like neutron stars (that are described by a Fermi gas), or Super-Chandrasekhar white dwarfs could be adequately modeled in this way. For such configurations it is evident that general relativistic effects as well as the inclusion of pressure anisotropy are unavoidable, and therefore its study could be carried on using the approach presented here.  Here too, care must be exercised with the fact that some of the physical phenomena present in such configurations (e.g. very strong magnetic fields) could break the spherical symmetry.

It is interesting to highlight that polytropes with $\gamma =1$ are used to model systems with non-degenerate isothermal cores which play an important role in the analysis of the Sch\"onberg-–Chandrasekhar limit. Any distribution of pressure anisotropy would affect the\break Sch\"onberg–-Chandrasekhar limit not only by affecting the structure of the polytrope, but also by the modifications of the virial theorem, introduced by the anisotropy. 

In the last section we have followed a similar strategy previously used to express the Tolman mass in terms of traces of the Riemann tensor (matter sector) and the only non--vanishing scalar associated to the electric part of the Weyl tensor in static and spherically symmetric space--times, $E$ 
(see \cite{lh20202complexity}, for example). In this sense, it is clear how the different parts 
affect the Tolman mass. For example, for conformally flat solutions the Weyl condition reads $E=0$ which leads to a geometric constraint which  allows to close the system. Similarly, here we have obtained expressions relating the scalar $\mathcal{K}$ with the traces of the Riemann tensor, the mass function $m$ (\ref{k:vs:m}) and the Tolman mass (\ref{eq:22:mtw:m}) and class one
solutions are obtained when the Karmarkar condition is fulfilled, namely $\mathcal{K}=0$. In this sense, expressions (\ref{k:vs:m}) and (\ref{eq:22:mtw:m}) may be useful to investigate the role played by Karmarkar embeding condition in our fluid distribution corresponding to the compact object. These aspects must be approached with the necessary care and detail in future works. 

We do not know if there are any real physical phenomena that can produce a distribution of local anisotropy similar to the one inherited through the metric Karmarkar condition described in our models. So, it could be interesting to be able to provide a physical meaningful picture for a source of the anisotropy that we have obtained (by means of magnetic field or charge distribution) of the type one could expect to find in some astrophysical settings. Although it is certainly an interesting question, the answer lies beyond the scope of this work. However, we believe that our approach provides a useful tool to carry out this and other investigations.

\section*{Acknowledgements}
The authors would like to acknowledge William Barreto for fruitful discussions on certain aspects of the numerical analysis.

\end{document}